\documentclass[12pt]{iopart}

\usepackage{graphicx}  
\begin{document}

\title[GRB polarization]{Polarization in the prompt emission of
gamma-ray bursts and their afterglows}

\author{Davide Lazzati}

\address{JILA, University of Colorado, 440 UCB, Boulder, CO
80309-0440, USA} \ead{lazzati@colorado.edu}
\begin{abstract}
Synchrotron is considered the dominant emission mechanism in the
production of gamma-ray burst photons in the prompt as well as in the
afterglow phase. Polarization is a characteristic feature of
synchrotron and its study can give a wealth of information on the
properties of the magnetic field and of the energy distribution in
gamma-ray burst jets. In this paper I will review the theory and
observations of gamma-ray bursts polarization. While the theory is
well established, observations have prove difficult to perform, due
to the weakness of the signal. The discriminating power of
polarization observations, however, cannot be overestimated.
\end{abstract}

\maketitle

\section{Introduction}
The ``standard'' model to interpret GRB emission implies dissipation
of bulk kinetic energy via collisionless
shocks~\cite{Rees92,Meszaros93,Piran99}. Magnetic fields are generated
in the process and highly relativistic electrons are accelerated in a
power-law distribution of energies (or Lorentz
factors)~\cite{Silva03,Hededal04}. In these conditions, radiation is
emitted through the synchrotron process, giving rise to a broad band
radiation source~\cite{Meszaros97,Piran99}.

Synchrotron radiation from a coherent magnetic field is known to be
polarized. As a consequence, theoretical and observational efforts
were produced to understand if and how much polarization should be
observed in the various phases of the GRB process. In this paper we
divide the GRB in two distinctive phases, the prompt and the afterglow.
For each phase we consider both theoretical predictions for linear
polarization (circular polarization is supposed to be very small,
especially in the afterglow phase~\cite{Matsumiya03}) and
observations. In both cases we will emphasize how the understanding of
the polarization properties of GRBs can shed light on other important
issues such as the release mechanism of GRB jets, the micro-physics of
collisionless shocks, the energy distribution of jets and the
properties of dust along the line of sight to the GRB.

\section{The afterglow phase}

Even if the afterglow phase follows the prompt phase in GRB
observations, it is worth and historically more correct to discuss the
polarization of the afterglow emission before that of the prompt
phase. This is due to the fact that the synchrotron nature of
afterglow photons is more firmly established and widely accepted,
compared to the highly debated origin of the prompt emission. For this
reason, theoretical predictions of linear polarization in the
afterglow phase appeared earlier than those related to the prompt
emission. As we shall see in the following, the prompt emission
polarization has, however, a larger potential to reveal the nature of
GRBs than that of the afterglow phase.

In order to discuss the polarization properties of synchrotron
radiation, we must first address the geometry of the magnetic field
that we expect in the afterglow. In the standard model, the afterglow
magnetic field is generated in the collisionless shock driven by the
fireball in the interstellar medium. Amplification of the interstellar
magnetic field is not sufficient to generate a magnetic field large
enough to explain the observed afterglow spectra and luminosities.

Numerical particle in cell (PIC) simulations of collisionless shocks
have become available only in recent years. The computing volume is 
still small and only several tens of skin depths can be followed (a
tiny fraction of the plasma volume supposed to be involved in the
afterglow production). Nevertheless, the two stream instability
(Weibel instability~\cite{medv99}) seem to be able to generate fields
close to equipartition that, within the explored volume, do not decay
due to diffusive dissipation~\cite{silva03,medv05,fred04}. Analogously
to compressed fields, such fields are tangled in the plane of the
shock but show a remarkable degree of alignment if observed
orthogonally to the normal of the shock plane~\cite{laing80}.

Linear polarization in GRB afterglows can be produced in at least two
ways by such fields. Gruzinov and Waxman~\cite{gru99} discussed the
possibility of creating coherent patches of magnetic field within the
visible area of the fireball by reorganizing the shock-produced field. 
They find that these patches can grow and linear polarization at the level of
\begin{equation}
P=\frac{70\%}{\sqrt{N}}\sim 10\%
\end{equation}
is expected, where $N\sim50$ is the expected number of visible
domain. The process is stochastic in nature and therefore a
straightforward prediction of the model is that polarization should be
subject to erratic variations of the position angle on timescales
$\delta{t}\sim{T}$, where $T$ is the time since the burst explosion.
This is due to the fact that any new magnetic domain that enters the
visible area can completely shift the polarization vector.
Observations of linear polarization in optical afterglows seem to fail
to detect such an erratic behavior, but see below for a more thorough
discussion.

\begin{figure}
\centerline{\includegraphics[height=11cm,width=11cm]{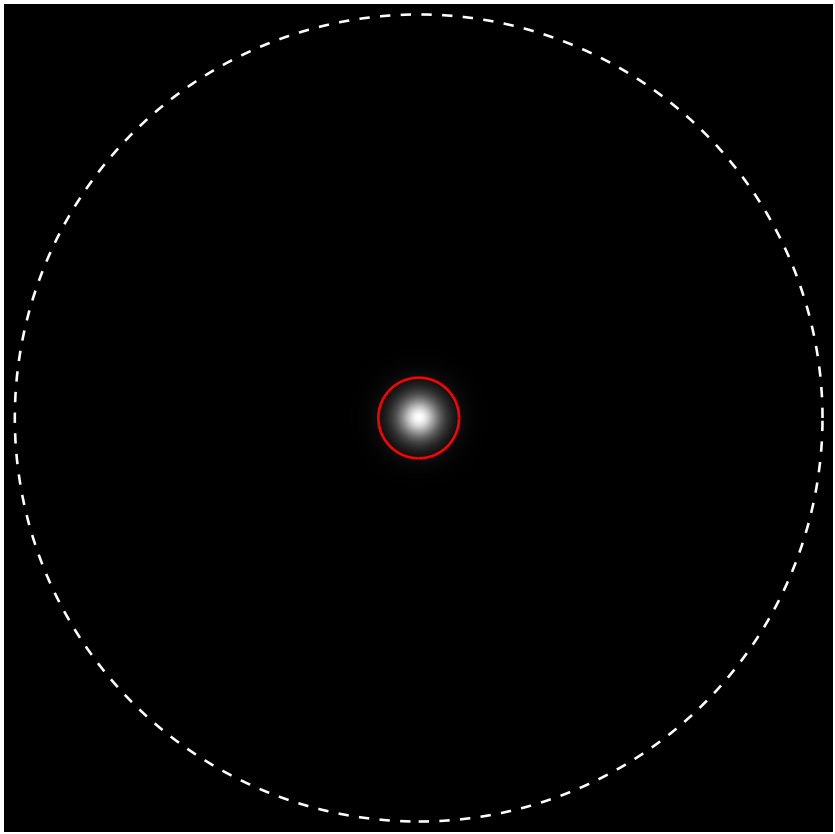}}
\caption{{Image of a relativistic fireball expanding with a constant
Lorentz factor $\Gamma=10$ and with a uniform luminosity (and flat
spectrum) in the comoving frame. The white dashed line shows the edge
of the fireball, while the red circle shows the region from which half
of the photons are observed (see Eq.~\ref{eq:aberr}).}
\label{fig:sphere}}
\end{figure}

The discovery of achromatic breaks in the afterglow decay of most GRBs
yielded to the conclusion that most GRBs are not spherical explosions
but beamed outflows. Polarization from a jet observed out of its
symmetry axis can be detected even if the magnetic field in the plane
of the shock is completely tangled~\cite{medv99,ghi99,sari99},
provided that it has the planar structure described above. To
understand this, let us first consider the effects of the aberration
of light in an expanding fireball. As a consequence of the
relativistic aberration of photons only a small fraction of the
fireball is visible. To be more precise, if the fireball is uniformly
luminous in the comoving frame\footnote{Note here that there is not a
single ``fireball comoving frame''. Since the fireball is expanding
radially, each point in the fireball defines its own comoving frame in
which every other part of the fireball is moving.}, half of the
photons received by the observer at infinity come from a small area
surrounding the line of sight of surface:
\begin{equation}
\Sigma=\pi\frac{R^2}{\Gamma^2}
\label{eq:aberr}
\end{equation}
where $R$ is the fireball radius and $\Gamma$ its Lorentz
factor. Figure~\ref{fig:sphere} shows a simulated image of a fireball
expanding at constant Lorentz factor $\Gamma=10$ and radiating
uniformly a flat featureless spectrum. The white dashed line shows the
edge of the fireball while the red circle shows the region in
Eq.~\ref{eq:aberr}. Consider now photons coming from the red circle in
the figure. In the comoving frame, they are produced in a direction
orthogonal to the shock. They see therefore a very coherent magnetic
field and are highly polarized in the direction parallel to the shock
speed. In the observer frame, they are boosted in a direction that
forms an angle $1/\Gamma$ with the shock speed and the polarization
vector (the electric vector) is rotated accordingly. Due to the shape
of the synchrotron spectrum and to the rapid variability of the
brightness, the observer at infinity sees a ring shaped
source~\cite{granot99} at the position of the red circle in the image.
The radiation from the ring is highly polarized in the local radial
direction. If the observer would be able to resolve the ring,
polarization would be observed. Unfortunately, the observer sees the
integrated radiation from the circle and different polarization
directions cancel out.

\begin{figure}
\centerline{\includegraphics[width=12cm]{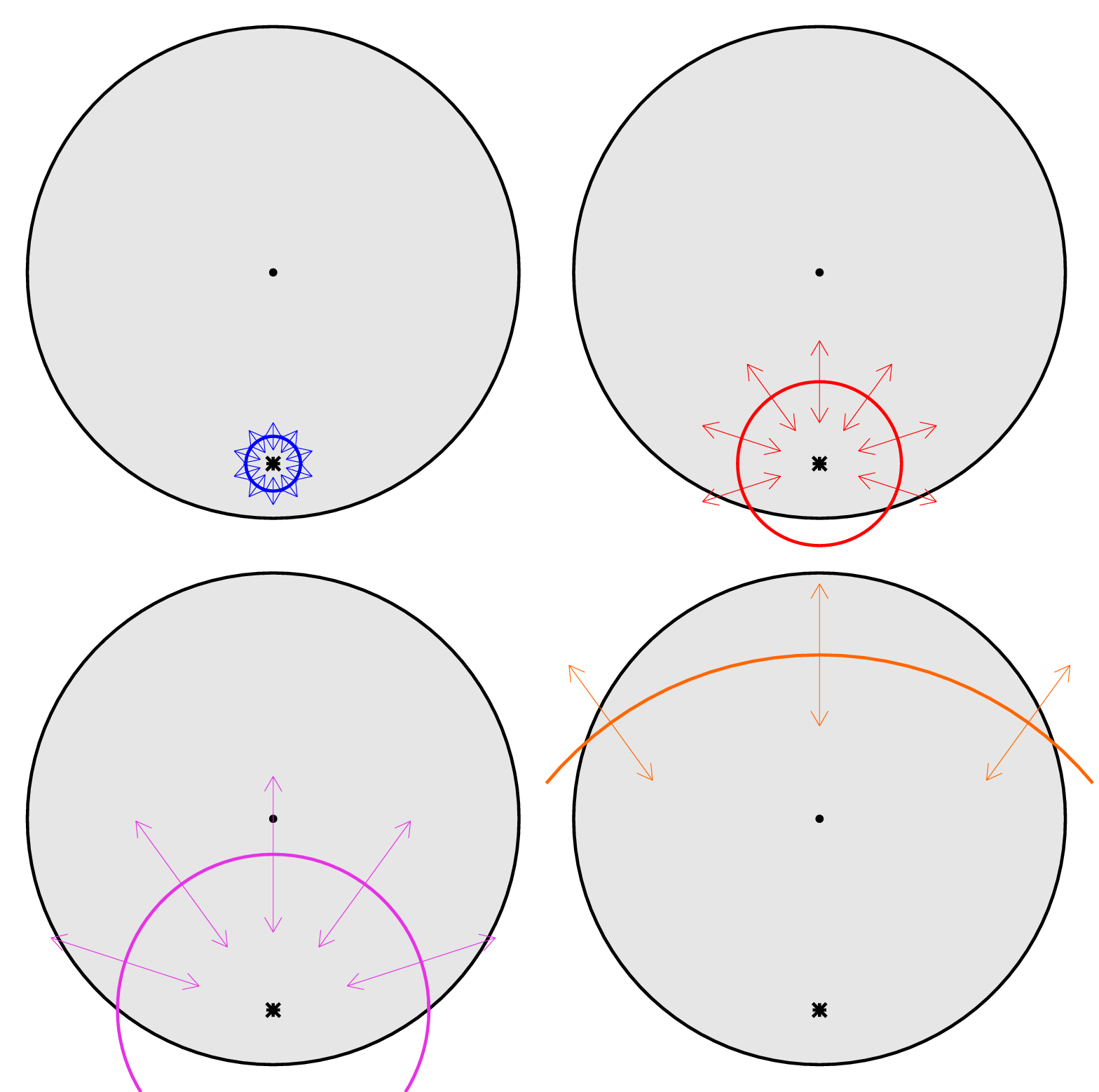}}
\caption{{Cartoon explaining the behavior of polarization from a
shock-generated magnetic field in a beamed fireball. The gray circle
shows the fireball seen face on. The asterisk in the lower center of
the circle shows the location of the line of sight. Time runs from
left to right and from top to bottom. The colored rings with arrows
show the location of the photon producing rings. The whole ring (blue)
is initially inside the fireball and no polarization is seen. At later
times, the lower part of the ring is lost and horizontal polarization
is detected (red). Finally (orange ring) only the top part of the ring
is seen and vertical polarization is detected. These two parts are
separated by a moment of vanishing polarization when only half of the
ring is visible (magenta).}\label{fig:rings}}
\end{figure}

An additional ingredient to this scenario is offered by the beaming of
jets. Consider an observer located slightly off-axis a beamed
outflow~\cite{ghi99,sari99}. He sees radiation coming from a circle
centered on the line of sight. At early stages, the radiation comes
from a small ring uniformly illuminated and no polarization is
observable. As time goes on and the fireball slows down, the size of
the ring increases and, eventually, one side of the ring hits the edge
of the jet. From this moment on, polarization will not cancel out
completely. Ghisellini \& Lazzati~\cite{ghi99} and Sari~\cite{sari99}
independently derived polarization curves for this scenario. A
schematic view of this process is shown in Fig.~\ref{fig:rings}. The
polarization is expected to be vanishing at very early times, since
the whole ring is inside the fireball when the Lorentz factor is very
high (upper left panel). As the fireball slows down, the ring from
which the radiation is mainly emitted grows until its lower part is
located outside of the fireball (upper right panel). In this
configuration, some vertical polarization is missing and therefore a
net horizontal polarization is observed. At some point, exactly half
of the ring is inside the fireball (the gray circle) and half of it is
outside. In this configuration (lower left panel) the two components
of polarization balance again and no net polarization is
observed. Finally, as the emission ring becomes larger and larger,
only a small upper portion in inside the fireball and all the
polarization becomes vertical.

\begin{figure}
\centerline{\includegraphics[width=12cm]{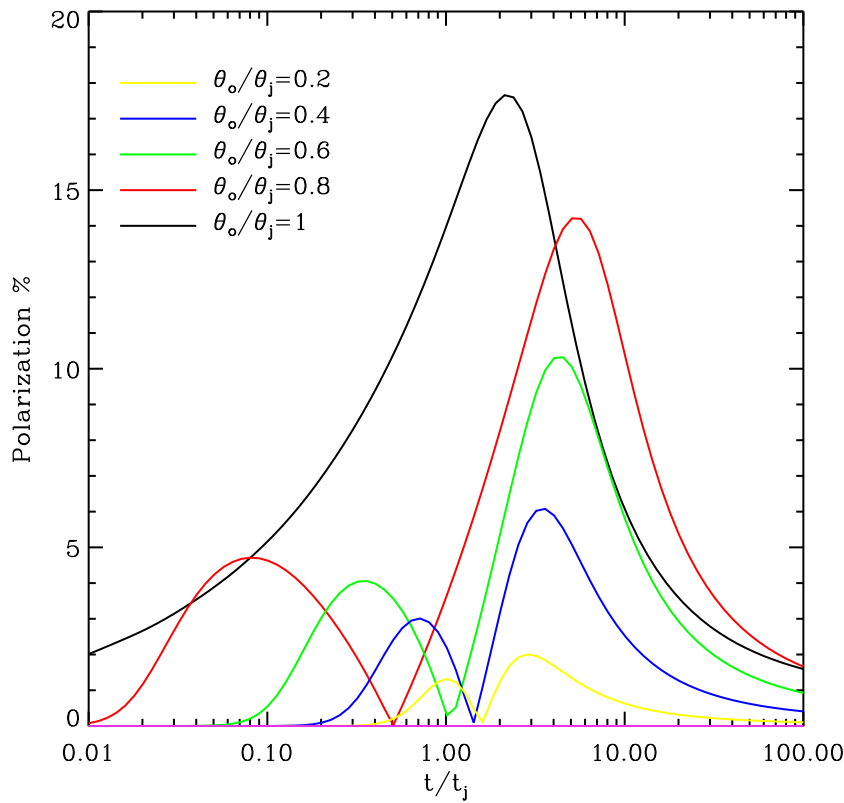}}
\caption{{Polarization curves from a top-hat jet with shock-generated
magnetic field. Different colors show polarization for different
viewing angles in units of the jet opening angle. All curves but the
one with $\theta_o=\theta_j$ have two polarization peaks. The
polarization angle in the two peaks is rotated by 90 degrees. The
black curve has only one peak that has the same orientation of the
second peak of the other curves.}\label{fig:pol}}
\end{figure}

This simple scenario can be complicated by the lateral expansion of
the jet~\cite{sari99}. Polarization curves have been computed, since
the model was originally proposed, with increased
refinement. Figure~\ref{fig:pol} shows the results obtained using the
code of Rossi et al.~\cite{Rossi04}. It was eventually shown that even
though the lateral expansion of the jet can modify the polarization
curve, it does not affect its general behavior and especially the
presence of the sudden rotation by $90^\circ$ of the position angle
approximately in concidence with the steepening in the light
curve~\cite{Granot03,Rossi04}.

After the discovery of polarization in
GRB990510~\cite{covino99,wijers99}, polarimetric observations in
search for the position angle rotation have been performed in a number
of bursts. Particularly good observations have been obtained for
GRB~021004~\cite{rol03}, GRB~020813~\cite{Lazzati04a} and
GRB~030329~\cite{Greiner03} (See Covino et al.~\cite{covino05} for a
complete review of polarization observations in GRB afterglows). For
the case of GRB~021004, it was initially claimed that the 90 degrees
rotation had been detected. However, it was subsequently shown that
the rotation had been observed at time earlier than
expected~\cite{Lazzati03}. The angle rotation is supposed to be
roughly associated to the time at which the jet geometry produces a
steepening in the afterglow light curve~\cite{rhoads99,sari99}. In the
case of GRB~021004 it was observed an order of magnitude earlier in
time~\cite{Lazzati03}. A possible explanation for the strange behavior
of the polarization angle of GRB~021004 is that its fireball is not
uniformly bright. The presence of prominent bumps in its light
curve~\cite{Lazzati02} is suggestive of such a case. If the fireball
is not uniformly bright, then the polarization component of the bright
spots dominates over the rest~\cite{Granot03,Lazzati03,nakar04}. The
polarization during flares can therefore be larger and with a random
orientation of the position angle. Additional modifications of the
polarization curve can be due to the propagation of the afterglow
photons in the ISM. Dust grains are dichroic and bi-birefringent and
imprint polarization and/or rotate the intrinsic one~\cite{Lazzati03}.
Local dust induced polarization is however easy to disentangle from
the prompt one with a suitable set of observations. It has a
well-known spectral dependence and is constant in time. High redshift
dust induced polarization is less well known. It is expected that
should be less severe (for a given amount of dust) since the grains
are expected to be smaller (the extinction curves are often analogous
to SMC templates). GRB afterglow spectropolarimetry is a great tool to
study high redshift dust. Unfortunately, so far no induced
polarization has been detected, probably due to the unavoidable bias
that associates induced polarization with extinction.

A more fortunate case is that of GRB~020813. This GRB has the
smoothest light curve measured so far, with stringent limits on its
variability (on top of the regular broken power-law
behavior)~\cite{Goro04}. Polarization measurements were performed
with good signal to noise before and after the jet break, an ideal
sample to check for the presence of the 90 degrees rotation of the
position angle. The modeling of the data showed that no rotation was
present, ruling out for this event a simple top-hat jet configuration
with shock-generated magnetic field~\cite{Lazzati04a}. Either the
structure of the jet or of its magnetic field have to be different.
Polarization from structured jets (with bright cores and dimmer wings)
was computed by Rossi et al.~\cite{Rossi04}. In this configuration,
for a shock generated magnetic field, the polarization position angle
is always toward the brightest part of the jet and therefore no
rotation of the position angle is expected. In addition, differently
from what observed in top-hat jets, the polarization has a maximum at
the jet break time rather than a minimum. Alternatively, it can be
assumed that the magnetic field has some degree of order, and is not
entirely shock generated. In this case the orientation of the magnetic
field dominates the position angle behavior that is, again,
constant. However (see also below for the case of the prompt
polarization) if the magnetic field is ordered, a large polarization
at early times is expected, contrary to any model in which the field
is shock-generated~\cite{Granot03,Lazzati04a}. The data of GRB~020813
are consistent with either scenario. In no case the polarization has
been observed at times early enough to allow to disentangle the effect
of the jet structure from that of the field orientation.

\begin{figure}
\centerline{\includegraphics[width=11cm]{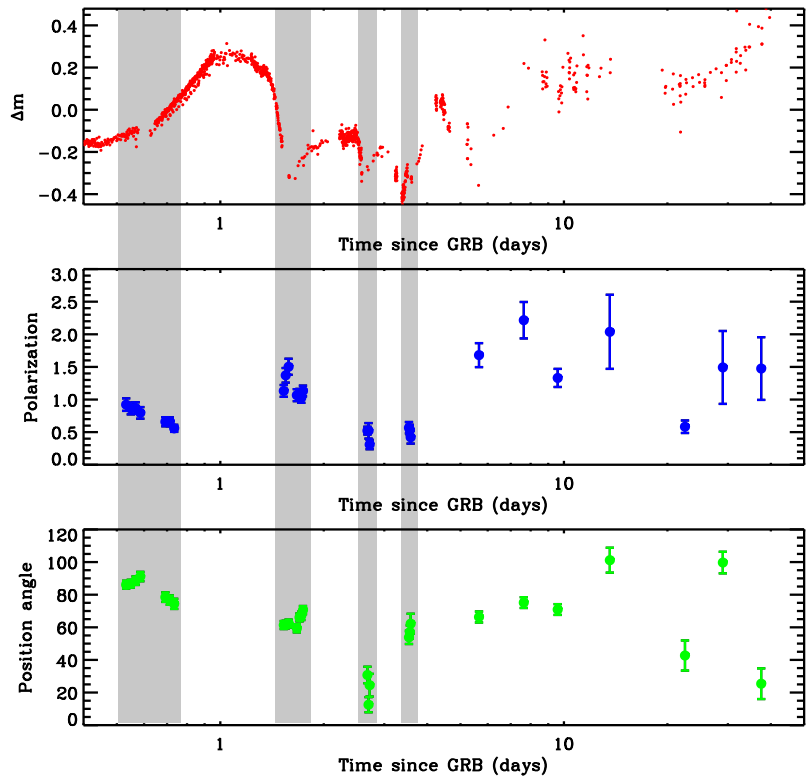}}
\caption{{The light curve, linear polarization and position angle of
GRB~030329. The top panel shows the R-band light curve~\cite{Lipkin04}
of GRB~030329 after the subtraction of a power-law model. This
emphasizes the presence of bumps and wiggles on top of the regular
power-law decay. The central panel shows the linear polarization data
and the bottom panel shows the relative position
angle~\cite{Greiner03}. Gray shadowed regions are overlaid to emphasize
the connection between the light curve bumps and the time at which
polarimetry was performed.}\label{fig:0329}}
\end{figure}

The afterglow with the best polarimetric sampling is, out of any
doubt, that of GRB~030329~\cite{Greiner03}. Linear polarimetry (and
even spectro-polarimetry) were performed at many times (see
Fig.~\ref{fig:0329}). Unfortunately, the light curve of GRB~030329 is
one of the least smooth, with variability overlaid on very short
timescales~\cite{Lipkin04}. The origin of this variability is not yet
clear, but it is out of doubt that it must involve a small fraction of
the fireball surface since the flares rise time is much shorter than
the curvature time scale~\cite{Lazzati02}. In such conditions it is
very hard to predict the behavior of polarization. The most
interesting feature in Fig.~\ref{fig:0329} is the change in the trend
of the position angle evolution during the first set of measurements
around $0.6$ days after the explosion. In this range of time the
light curve is smooth, the polarization is uniformly decreasing, yet
the position angle trend has a break. Such a behavior is not easy to
interpret in any of the theoretical framework outlined above. Being a
single case it is possible to argue that is a coincidence, and indeed
it is possible that a non-uniformly bright fireball produces a smooth
light curve, a smooth polarization curve and yet has random position
angle variations~\cite{nakar04}.

In summary, the theory of afterglow polarization is well developed,
and is ahead of observations. Crucial observations at very early
stages and a dense polarization sampling of a smooth afterglow
light curve are still to be performed. The only conclusion we can draw,
from the modeling of GRB~020813~\cite{Lazzati04a}, is that the
simplest jet model fails to explain the observations.

\section{The prompt phase}

The prompt phase of GRBs is the brightest phase of the phenomenon and
is characterized by strong variability and typical photon energies in
the hundreds of keV range. These features are thought to be due to the
relaxation of instabilities that are injected in the base of the jet,
are advected by the flow and reach a catastrophic point at about
$R_{\rm{IS}}\simeq10^{14}$~cm from the base of the jet.

The prompt emission is therefore thought to originate from the jet
material, interacting with itself (from which the name of ``internal
shocks''). To date, it has been impossible to deconvolve the pulse
properties in order to derive physical parameters on the jet
properties in the prompt phase. As a consequence, hydrodynamic jet
models (e.g., \cite{Lazzati05}) coexist with fully magnetic models
(e.g., \cite{Lyutikov03}) and little is known about the jet launching
process, its composition and its dynamics. The synchrotron radiation
mechanism itself is highly
debated~\cite{Lyutikov03,Ghisellini00,Lazzati00} as well as the
dissipation method (usually identified with relativistic
shocks)~\cite{Lazzati99}.

\subsection{Polarization in the prompt phase}

\begin{figure}
\centerline{\includegraphics[height=10cm,width=11cm]{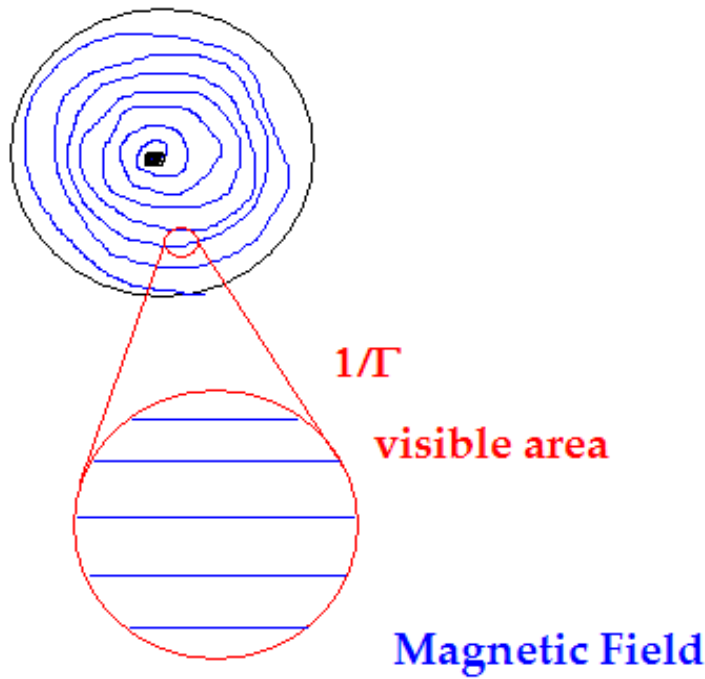}}
\caption{{Effect of the relativistic beaming on the geometry of the
observed magnetic field. Since only a small area of the fireball can
be observed, a toroidal magnetic field permeating the jet can have a
large degree of coherence in the visible area.}
\label{fig:b1}}
\end{figure}

For the reasons highlighted above, any additional information on the
prompt phase is extremely important and precious. At the end of 2002
it was announced that the prompt emission of GRB~021206 was linearly
polarized to the astonishingly high level of $80\%$
\cite{Coburn03}. Even though the detection turned out to be much less
robust than initially claimed (see below for a discussion), it
stimulated a very intense and productive theoretical effort.

A first result was the realization that relativistic effects in a
spherical (or conical) fireball reduce the observed degree of
polarization~\cite{Lyutikov03,Granot03a,Lazzati04}. This effect is due
to the aberration of light. As discussed above, due to the light
aberration only a small portion of the fireball is visible to the
observer (see Fig.~\ref{fig:sphere}). This has two effects. On the one
hand (see Fig.~\ref{fig:b1}) the magnetic field appears very well
aligned even if the overall structure is more complex. For example, a
non-relativistic jet with a toroidal magnetic field seen head-on (or
slightly off-axis) is non-polarized. A relativistic jet, instead, has
a large degree of polarization, close to the maximum polarization
achievable from synchrotron. 

A second effect, however, counterbalance this. Consider the zoomed
part in Fig.~\ref{fig:b1}, which represents the visible part of the
fireball. The regions of that area close to its edge move with a speed
that makes an angle with respect to the line of sight. Do to
relativistic aberration, the polarization produced by the edge of the
area will be rotated~\cite{Lyutikov03}. This causes a net decrement of
the observed polarization. This effect affects both the intrinsic and
geometric models for prompt polarization.

\subsubsection{Intrinsic models}

Consider now a structure like the one in Figure~\ref{fig:b1}. The
fireball jet, that is observed face on in the upper part of the
figure, is permeated by a toroidal magnetic field. This is a likely
configuration if a sizable magnetic field is advected by the jet from
its base. Due to the conservation of magnetic flux, the radial
component of the magnetic field will decrease as
$B_\parallel\propto{}R^{-2}$, faster than the orthogonal component
$B_\perp\propto R^{-1}$.  At large radii (when $\gamma$-ray photons
are produced the radius has increased by 5 orders of magnitude or
more) only the orthogonal component survives. The bottom part of the
figure shows how the magnetic field would appear to an observer that
lies away from the jet axis (the singular point of a toroidal magnetic
field). Since only a small area is visible, the field is observed as
completely parallel.  

\begin{figure}
\centerline{
\parbox{0.48\textwidth}{
\includegraphics[width=0.47\textwidth,height=0.47\textwidth]{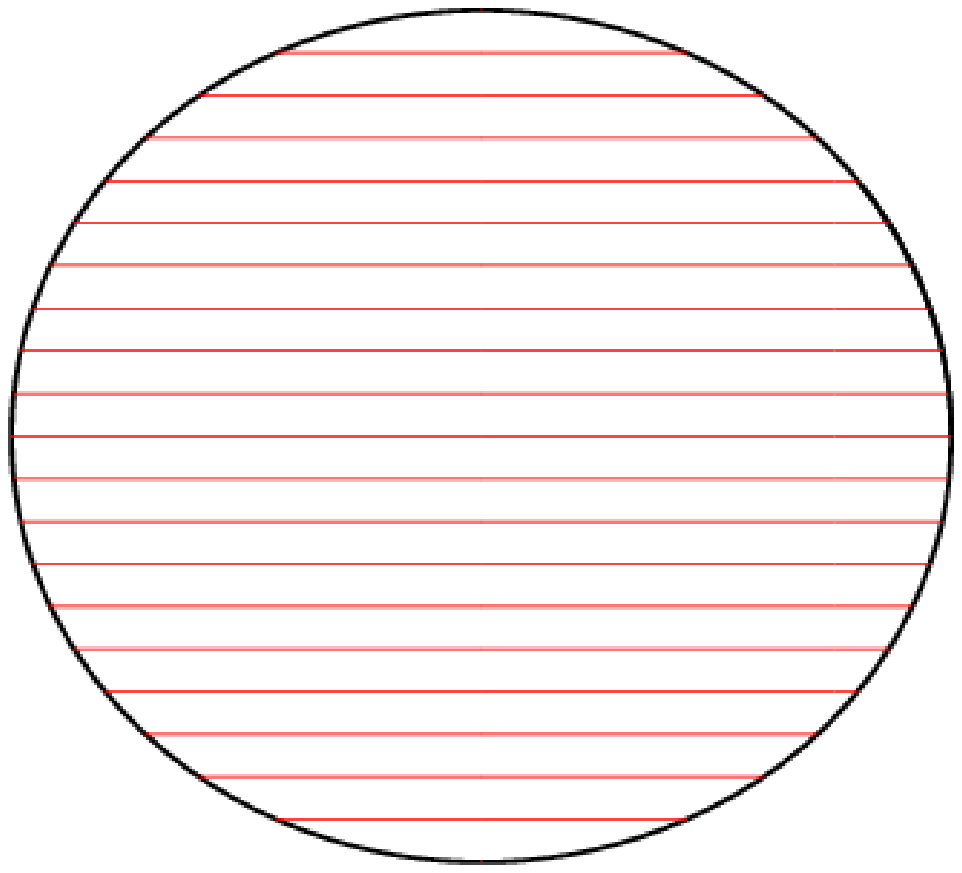}}
\hspace{0.03\textwidth}
\parbox{0.48\textwidth}{
\includegraphics[width=0.47\textwidth,height=0.47\textwidth]{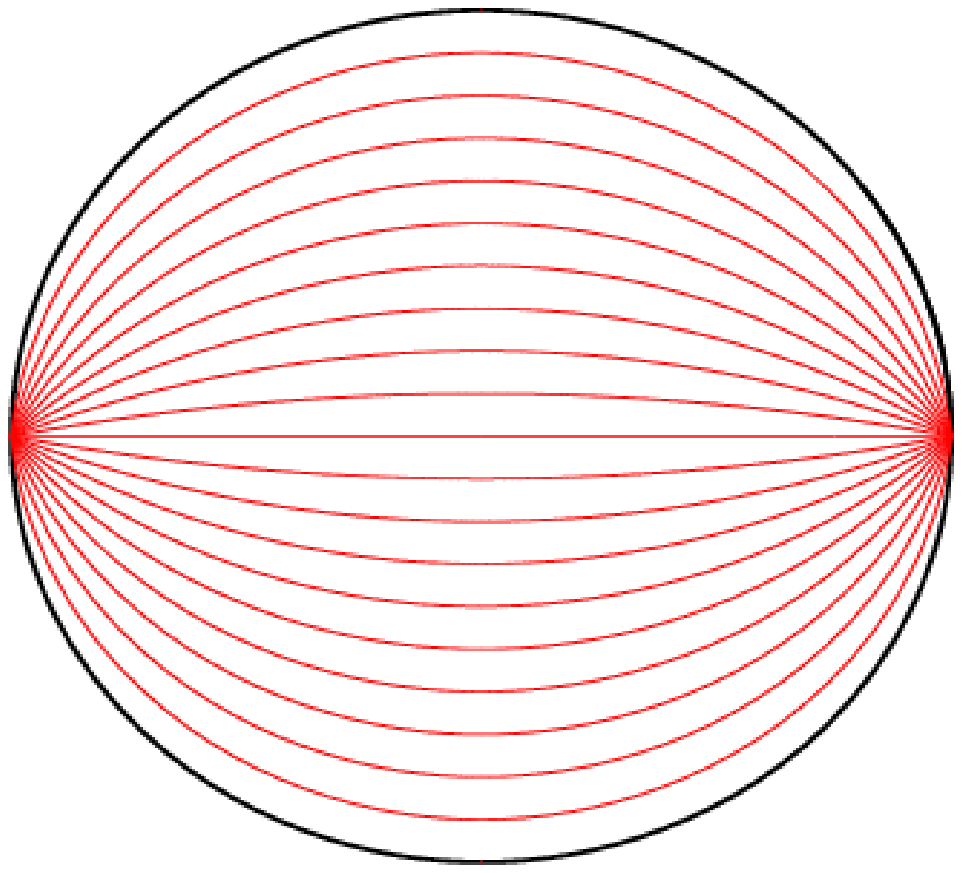}}
}
\caption{{Apparent direction of the magnetic field as derived by
assuming it is orthogonal to the local polarization direction. The
left panel shows the B lines without taking into account the
relativistic aberration of photons, while the right one shows the
correct image~\cite{Lyutikov03,Granot03a,Lazzati04}}\label{fig:babe}}
\end{figure}

This in not, however, the only effect of relativistic aberration that
plays a role in the polarization of GRB prompt
emission. Figure~\ref{fig:babe} shows the observed magnetic field
geometry, as reconstructed by assuming it is orthogonal to the local
direction of the linear polarization. The left panel is analogous to
the zoomed area in Fig.~\ref{fig:b1}, where all the field is perfectly
aligned. However, the Lorentz boost that connects the different parts
of the fireball to the observer is different in the center of the area
from its edges. In particular, the boost at the edges makes an angle
with the line of sight direction. This implies an angle rotation of
the electric vector and therefore of
polarization~\cite{Lyutikov03,Granot03a}. The right panel of
Fig.~\ref{fig:babe} shows the apparent magnetic field force lines
after this effect has been taken into account.

\begin{figure}
\centerline{\includegraphics[width=0.6\textwidth]{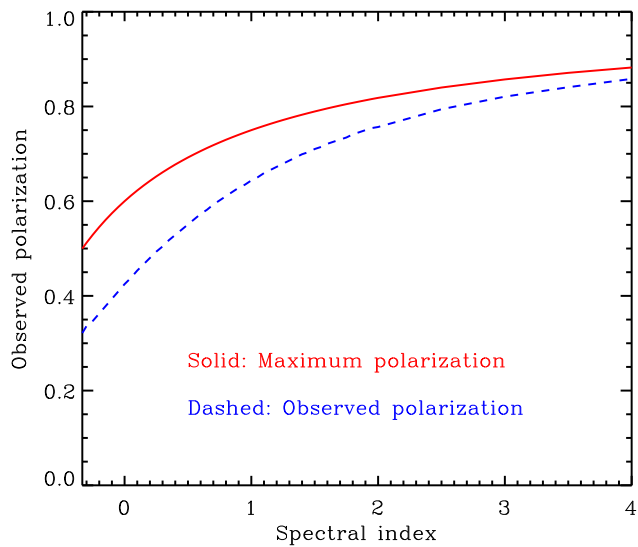}}
\caption{{Resulting polarization of the prompt emission of GRBs from a
completely aligned magnetic field as a function of the power-law
spectral index. The red line shows the theoretical maximum
polarization while the blue dashed line shows the polarization that is
observed after light aberration has been taken into
account.}\label{fig:curv}}
\end{figure}

An additional ingredient is necessary in order to compute the amount
of net polarization resulting after the aberration is taken into
account. The local brightness of the fireball depends on the spectral
slope since $I(\nu)=\delta^{3+\alpha}I^\prime(\nu^\prime)$, where
$\delta$ is the Doppler factor and $I(\nu)\propto\nu^{-\alpha}$. As a
result, a flatter spectrum will have more flux in the edges of the
visible area, and the reduction of polarization will be
larger~\cite{Lyutikov03,Granot03a}. Fig.~\ref{fig:curv} shows the
resulting polarization for a power-law spectrum as a function of the
spectral index.

\begin{figure}
\centerline{\includegraphics[width=0.6\textwidth]{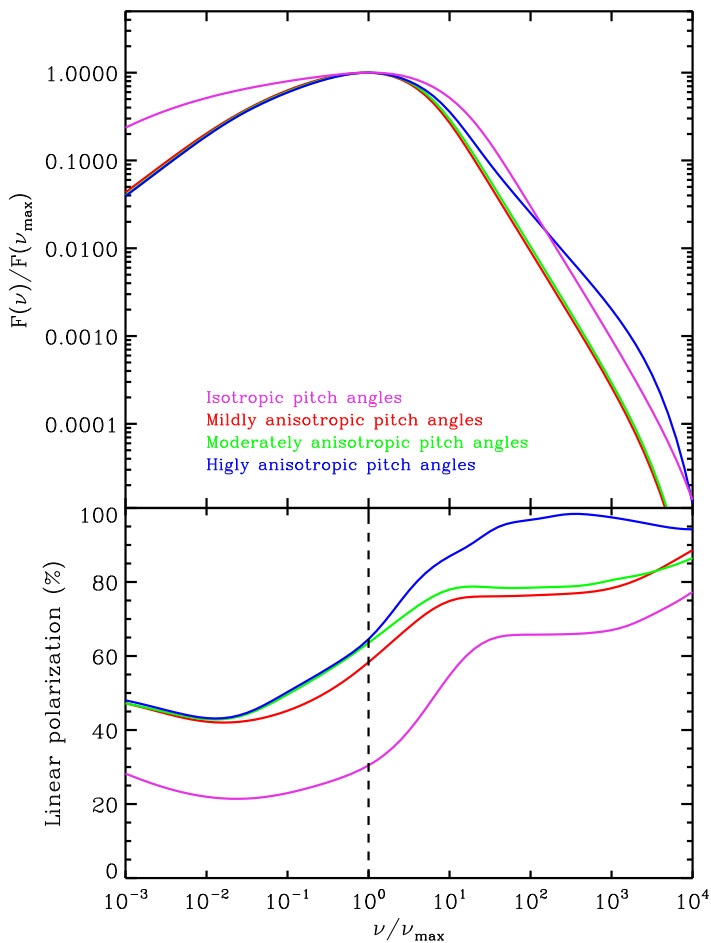}}
\caption{{Spectra and polarization of an anisotropic distribution of
electrons in a relativistic fireball.}\label{fig:pitch}}
\end{figure}

The observations of GRB~021206\footnote{In the following discussion we
will assume the original measurement of polarization is
correct~\cite{Coburn03}. The result has been however severely
criticized and may not be correct~\cite{rutledge04,wigger04}.} showed a
very high level of polarization only marginally consistent with the
values shown in Fig.~\ref{fig:curv}. It is possible to envisage
scenarios in which the observed polarization could be larger. For
example, the intensity distribution of the fireball and the maximum
polarization are modified if the pitch angle distribution of the
electrons is not isotropic but rather biased toward the orthogonal
direction. Some acceleration mechanisms produce such an anisotropic
distribution. Figure~\ref{fig:pitch} shows the spectrum and
polarization of relativistic electrons with an anisotropic pitch angle
distribution. The more anisotropic is the distribution, the larger is
the net polarization observed.

\subsubsection{Geometric models}

In alternative to the models described above, that we called
intrinsic, high polarization in the prompt phase can be produced by
another class of models, the so-called geometric
models~\cite{waxman03,Gruzinov99,Lazzati04}. In these models, the
polarization would be null in a normal geometric configuration, but is
enhanced by the simultaneous realization of two conditions. First, the
fireball must be very narrow, with an opening angle
$\theta_j\sim1/\Gamma$ where $\Gamma$ is the fireball Lorentz
factor. This condition ensures that the Doppler factor for the whole
fireball is the same. Secondly, the viewing angle must be such that
$\theta_o\sim2\theta_j$. This condition ensures that the photons that
the observer sees are mostly produced parallel to the fireball surface
in the comoving frame.

\begin{figure}
\centerline{\includegraphics[width=0.7\textwidth]{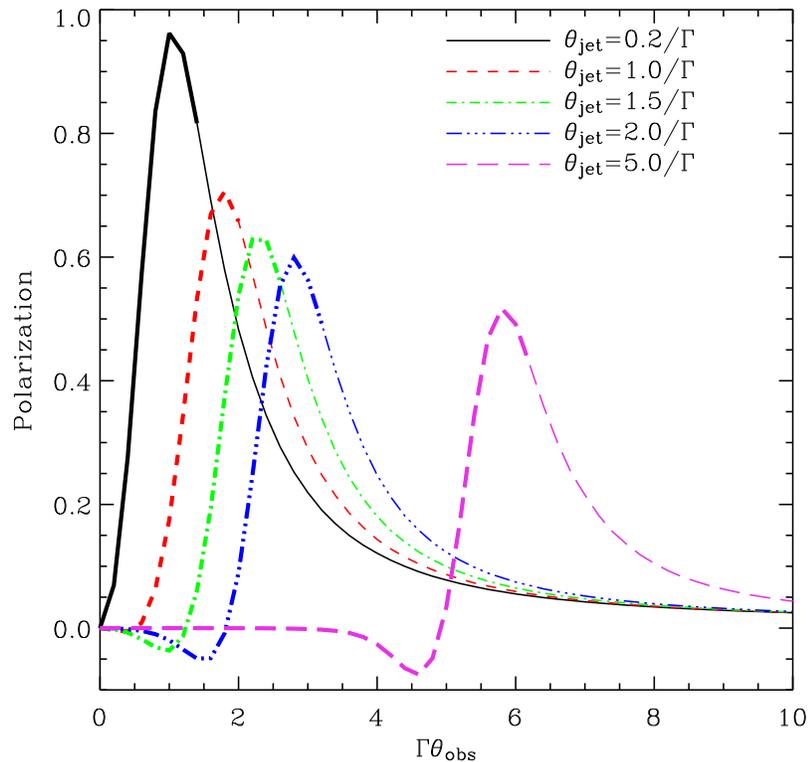}}
\caption{{Polarization curves as a function of the observing angle in
unit of the relativistic beaming angle. Different curves (see inset)
shows polarization from fireballs with different opening angles. The
black is the most narrow, while the magenta is the
widest.}\label{fig:cd}}
\end{figure}

If such geometric conditions are satisfied, high polarization of the
prompt emission can be achieved for a shock generated
field~\cite{Nakar03} and even if the prompt emission is due to bulk
inverse Comptonization of field photons~\cite{Lazzati00,Ghisellini00}.
In the synchrotron case, the polarization is analogous to the
afterglow case, but the net result is enhanced by the particular
geometric condition, that ensures no cancellation takes place due to
the different orientations of the field. In the Compton drag case, the
geometric conditions ensure that the observer at infinity detects
primarily photons that had, in the electron comoving frame, a
scattering of 90 degrees. IC scattering polarize photons as
\begin{equation}
\Pi=\frac{\sin^2\theta}{1+\cos^2\theta}
\end{equation}
where $\theta$ is the scattering angle. Polarization can therefore be
complete.  Figure~\ref{fig:cd} shows the resulting polarization as
a function of the two geometric conditions outlined
above. Polarization can be substantial, but a sizable reduction is
always observed even in the more optimistic cases.

The possibility of realizing such strict conditions has been
questioned.  A-posteriori statistics is always difficult to
compute. The simultaneous realization of the two geometric conditions
is indeed hard, but some circumstances help. The most important one is
that GRB~021206 was very bright, even for a moderate redshift. If we
assume it lied on the Frail and Amati
correlations~\cite{Frail01,Amati02}, it should have had a very small
opening angle. This eases largely the constraints on the geometric
conditions.

\section{In between: polarization of the optical flash}

The claimed polarization of GRB~021206~\cite{Coburn03} sparked intense
theoretic and observational activities. It showed that high-energy
polarization is an extremely sensitive tool to explore GRB jet
structures and composition. However, it also showed that performing
the measurements is very difficult, and a dedicated experiment should
be designed. A shortcut may be the observation of polarization from
the optical flash~\cite{Granot03}.

When the fireball impacts on the external medium to produce the
external shock, a transient shock is produced inside the fireball
itself. This is usually called ``reverse shock''. The emission from
this shock lasts for a short time (the time the shock takes to cross
the fireball) and is peaked in the optical. From the polarization
point of view, this emission could be similar to the prompt one more
than to the afterglow one. The lack of bright optical flashes in the
Swift era has so far prevented the observation of very early time
polarization. As for the afterglow case, the theory is ready to
confront the observations.

\section*{Acknowledgments}

This review is the outcome of many years of work, shared with many
collaborators. I'd like to thank in particular G. Ghisellini,
A. Celotti, M. J. Rees, E. Rossi, and S. Covino for their invaluable
help. This work was supported by NSF grant AST-0307502 and NASA
Astrophysical Theory Grant NAG5-12035.


\section*{References}
\begin{thebibliography}{99}
\bibitem{Rees92} Rees, M.~J., \& Meszaros, P.\ 1992, MNRAS, 258, 41P
\bibitem{Meszaros93} Meszaros, P., \& Rees, M.~J.\ 1993, ApJ, 405, 278
\bibitem{Piran99} Piran, T.\ 1999, Physics Reports, 314, 575
\bibitem{Silva03} Silva, L.~O, \etal, M.~V.\ 2003, ApJ, 596, L121
\bibitem{Hededal04} Hededal, C.~B., \etal, 2004, ApJ, 617, L107
\bibitem{Meszaros97} Meszaros, P., \& Rees, M.~J.\ 1997, ApJ, 476, 232
\bibitem{Matsumiya03} Matsumiya, M., \& Ioka, K.\ 2003, ApJ, 595, L25
\bibitem{medv99} Medvedev, M.~V. \& Loeb, A., 1999, ApJ, 526, 697
\bibitem{silva03} Silva, L.~O., Fonseca, R.~A., Tonge, J.~W., Dawson,
  J.~M., Mori, W.~B., \& Medvedev, M.~V.\ 2003, ApJ, 596, L121
\bibitem{medv05} Medvedev, M.~V., Fiore, M., Fonseca, R.~A., Silva,
  L.~O., \& Mori, W.~B.\ 2005, ApJ, 618, L75
\bibitem{fred04} Frederiksen, J.~T., Hededal, C.~B., Haugb{\o}lle, T.,
  \& Nordlund, {\AA}.\ 2004, ApJ, 608, L13
\bibitem{laing80} Laing, R.~A.\ 1980, MNRAS, 193, 439
\bibitem{gru99} Gruzinov, A., \& Waxman, E.\ 1999, ApJ, 511, 852
\bibitem{ghi99} Ghisellini, G., \& Lazzati, D.\ 1999, MNRAS, 309, L7
\bibitem{sari99} Sari, R.\ 1999, ApJ, 524, L43\bibitem{Lipkin04} Lipkin Y.~M., \etal, 2004, ApJ, 606, 381

\bibitem{granot99} Granot, J., Piran, T., \& Sari, R.\ 1999, ApJ,
  513, 679
\bibitem{Rossi04} Rossi E.~M., Lazzati D., Salmonson J.~D., Ghisellini
  G., 2004, MNRAS, 354, 86
\bibitem{Granot03} Granot J., K{\"o}nigl A., 2003, ApJ, 594, L83
\bibitem{covino99} Covino S., \etal, 1999, A\&A, 348, L1
\bibitem{wijers99} Wijers R.~A.~M.~J., \etal, 1999, ApJ, 523, L33
\bibitem{rol03} Rol E., \etal, 2003, A\&A, 405, L23
\bibitem{Lazzati04a} Lazzati D., \etal, 2004, A\&A, 422, 121
\bibitem{Greiner03} Greiner J., \etal, 2003, Nature, 426, 157
\bibitem{covino05} Covino S., Rossi E., Lazzati D., Malesani D.,
  Ghisellini G., 2005, AIPC, 797, 144
\bibitem{Lazzati03} Lazzati D., et al., 2003, A\&A, 410, 823
\bibitem{Lazzati02} Lazzati D., Rossi E., Covino S., Ghisellini G.,
  Malesani D., 2002, A\&A, 396, L5
\bibitem{rhoads99} Rhoads J.~E., 1999, ApJ, 525, 737
\bibitem{nakar04} Nakar E., Oren Y., 2004, ApJ, 602, L97
\bibitem{Goro04} Gorosabel J., \etal, 2004, A\&A, 422, 113
\bibitem{Lipkin04} Lipkin Y.~M., \etal, 2004, ApJ, 606, 381
\bibitem{Lazzati05} Lazzati, D., \& Begelman, M.~C.\ 2005, ApJ, 629,
  903
\bibitem{Lyutikov03} Lyutikov, M., Pariev, V.~I., \& Blandford, R.~D.\
  2003, ApJ, 597, 998
\bibitem{Ghisellini00} Ghisellini, G., Celotti, A., \& Lazzati, D.\
  2000, MNRAS, 313, L1
\bibitem{Lazzati00} Lazzati, D., Ghisellini, G., Celotti, A., \& Rees,
  M.~J.\ 2000, ApJ, 529, L17
\bibitem{Lazzati99} Lazzati, D., Ghisellini, G., \& Celotti, A.\ 1999,
  MNRAS, 309, L13
\bibitem{Coburn03} Coburn, W., \& Boggs, S.~E.\ 2003, Nature, 423, 415
\bibitem{Granot03a} Granot, J.\ 2003, ApJ, 596, L17
\bibitem{Lazzati04} Lazzati, D., \etal, 2004, MNRAS, 347, L1
\bibitem{rutledge04} Rutledge R.~E., Fox D.~B., 2004, MNRAS, 350, 1288
\bibitem{wigger04} Wigger C., Hajdas W., Arzner K., G{\"u}del M.,
  Zehnder A., 2004, ApJ, 613, 1088
\bibitem{waxman03} Waxman E., 2003, Nature, 423, 388
\bibitem{Gruzinov99} Gruzinov, A., \& Waxman, E.\ 1999, ApJ, 511, 852
\bibitem{Nakar03} Nakar E., Piran T., Waxman E., 2003, JCAP, 10, 5
\bibitem{Frail01} Frail D.~A., \etal, 2001, ApJ, 562, L55
\bibitem{Amati02} Amati L., \etal, 2002, A\&A, 390, 81
\end{thebibliography}
\end{document}